\documentclass[12pt]{article}
\usepackage{amsmath}
\usepackage{amstext,amssymb,amsfonts,color} 
\usepackage[english, USenglish]{babel}
\usepackage{graphicx}

\advance\voffset by -1.5cm
\advance\hoffset by -1.25cm
\definecolor{darkgreen}{rgb}{0,0.6,0}
\def\thefootnote{\fnsymbol{footnote}}

\def\be{\begin{equation}}
\def\ee{\end{equation}}
\def\ba{\begin{eqnarray}}
\def\ea{\end{eqnarray}}

\newcommand{\ZZ}{\mathbb{Z}}

\newcommand{\thetaR}[2]{\theta \!\begin{bmatrix} #1 \\ #2 \end{bmatrix}\!}

\newtheorem{thm}{Theorem}[section]

\newtheorem{prop}[thm]{Proposition}

\begin{document}


\thispagestyle{empty}
\renewcommand{\thefootnote}{\fnsymbol{footnote}}

{\hfill \parbox{2.45cm}{
 DESY 12-189 \\ 
}}

\bigskip

\begin{center} \noindent \Large \bf
Spin-\boldmath{$k/2$}-spin-\boldmath{$k/2$}  \boldmath{$SU(2)$}  two-point functions\\
 on the torus
\end{center}

\bigskip\bigskip\bigskip

\centerline{ \normalsize \bf 
Ingo Kirsch$^a$ 
  \footnote[1]{\noindent \tt email: ingo.kirsch@desy.de} 
and Piotr Kucharski$^b$ 
  \footnote[2]{\noindent \tt email: pjfkucharski@gmail.com} }

\bigskip

\centerline{\it ${}^a$ DESY Hamburg, Theory Group,}
\centerline{\it Notkestrasse 85, D-22607 Hamburg, Germany}
\vspace{0.3cm}
\centerline{\it ${}^b$ Institute of Theoretical Physics, Faculty of Physics, University of Warsaw}
\centerline{\it ul. Ho\.{z}a 69, 00-681 Warsaw, Poland}

\bigskip\bigskip

\bigskip\bigskip

\renewcommand{\thefootnote}{\arabic{footnote}}

\centerline{\bf \small Abstract}
\medskip

{\small \noindent We discuss a class of two-point functions on the
  torus of primary operators in the $SU(2)$ Wess-Zumino-Witten model
  at integer level $k$. In particular, we construct an explicit
  expression for the current blocks of the
  spin-$\frac{k}{2}$-spin-$\frac{k}{2}$ torus two-point functions for all $k$.  We
  first examine the factorization limits of the proposed current
  blocks and test their monodromy properties. We then prove that the
  current blocks solve the corresponding Knizhnik-Zamolodchikov-like
  differential equations using the method of Mathur, Mukhi and Sen.}

\newpage
\setcounter{tocdepth}{2}
\tableofcontents

\setcounter{equation}{0}
\section{Introduction}

The $SU(2)$ Wess-Zumino-Witten (WZW) model is an integral part of
string theory on $AdS_3 \times S^3$. In this string theory, 
worldsheet correlators were computed on the sphere
and agreement with the correlators in the dual conformal field theory
was found~\cite{Gaberdiel}. The computation was done in a particular large $N$ limit
in which higher-genus contributions are  suppressed. In order to
leave the large $N$ limit  in AdS/CFT \cite{Maldacena}, 
it would be desirable to repeat such computations on higher-genus
surfaces. Unfortunately, while correlators in the $SU(2)$ WZW model have been much
studied on the sphere \cite{FZ, Dotsenko}, not much is known at higher
genera (see however \cite{B1}-\cite{Gawedzki2} for general progress).
This makes it difficult to go beyond large~$N$ and
motivates our renewed interest in the $SU(2)$ WZW model on Riemann
surfaces with genus $g \geq 1$.

A few correlators of $SU(2)$ primary fields are known on the torus 
($g=1$). The zero-point functions, {\em i.e.}\ the characters, were  
determined in \cite{KP}.
 The one-point functions \cite{Suzuki} vanish as they are
not singlets under $SU(2)$.  The two-point functions were determined
at level $k=1, 2$ by the identification of the $SU(2)$ WZW model with
free field theories~\cite{Schnitzer}, by solving appropriate
differential equations \cite{MMS, MMS2, MM}, and by pinching genus-2
characters \cite{BKM, Choi}.  Not so much is known for levels $k \geq
3$ at which the $SU(2)$ WZW model becomes interacting. An integral
representation of the two-point spin-$\frac{1}{2}$-spin-$\frac{1}{2}$
current blocks on the torus  was found in \cite{Jayaraman:1989tu} (for all $k$),
using a Coulomb gas approach, and in \cite{Smyrnakis}, using the free
field representation developed by Bernard and
Felder~\cite{BernardFelder}.  In principle, these methods can be used
to find expressions for the other classes of torus two-point functions
but may be cumbersome to apply. Other correlators on the torus and
related work can be found in \cite{Dolan, Gannon}.

In this paper we find a relatively simple expression for the
spin-$\frac{k}{2}$-spin-$\frac{k}{2}$ torus two-point functions
$\langle \Phi_{k/2}(z)\Phi_{k/2}(0)\rangle$, where $\Phi_{k/2}$ is a
level-$k$ primary with maximal spin $j=\frac{k}{2}$ and conformal
weight $h=\frac{k}{4}$.  A~special feature of these two-point
functions is that they only have the identity running in their
intermediate channel. In the limit, when the two insertion points are
close together, their current blocks therefore factorize into the
two-point function on the sphere and a level-$k$ character.  This
suggests to write the current blocks~as
\begin{align}
f_l \sim \frac{\chi_l(z,\tau)}{E(z,\tau)^{2h}} \,,\label{ansatz}
\qquad (l=0,..., \textstyle \frac{k}{2})\,,
\end{align}
where $E(z,\tau)$ is the torus prime form, and $\chi_l(z,\tau)$ is an
extension of the character $\chi_l(\tau)=\lim_{z\rightarrow 0}
\chi_l(z,\tau)$ to non-zero values of the distance $z$ of the two
insertions.  The non-trivial task is to find a function
$\chi_l(z,\tau)$ which has the correct monodromy properties with
respect to shifts $z\rightarrow z+1$ and $z\rightarrow z+\tau$.
Fortunately, in the spin-$\frac{k}{2}$ case, the expression
$\chi_l(z,\tau)$ provided by Kac and Peterson \cite{KP} already has
the correct monodromy properties, and the $f_l$'s given by
(\ref{ansatz}) become the current blocks of $\langle
\Phi_{k/2}(z)\Phi_{k/2}(0)\rangle$.  This will be proven using the
differential equation technique developed by Mathur, Mukhi and Sen
\cite{MMS, MMS2} (see also \cite{Kiritsis}).

The paper is organized as follows. In section~\ref{sec2}, we review
some basic prerequisites and state our conjecture for the
spin-$\frac{k}{2}$-spin-$\frac{k}{2}$ torus two-point functions.  In
section~\ref{sec3}, we examine the factorization limits of the
corresponding current blocks and study their mono\-dromy properties.
In section~\ref{sec4}, we eventually prove our conjecture using the
differential equation technique of \cite{MMS, MMS2}. In section~\ref{sec5},
we conclude with a few comments on other $SU(2)$ torus two-point 
functions.

\setcounter{equation}{0}
\section{The spin-$\frac{k}{2}$-spin-$\frac{k}{2}$ two-point function on the torus}\label{sec2}

Consider the $SU(2)$ WZW model at integer level $k$. It is invariant
under the combined action of the Virasoro and affine ${su}(2)$
algebra.  The primary fields $\Phi_j$ have the conformal weights
\begin{align}
h_j=\frac{j(j+1)}{k+2}\,, \qquad j=0, ...,\textstyle\frac{k}{2} \,,
\end{align}
and satisfy
\begin{align}
J_n^a |j\rangle = 0 \,, \quad n>0\,,\qquad
J_0^+  |j\rangle = 0 \,,\qquad
J^0_0 |j\rangle = j |j\rangle \,,
\end{align} 
where $J^a_n$ ($a={\pm,0}$) are the modes of the holomorphic $SU(2)$
currents. The central charge is $c=\frac{3k}{k+2}$.

\begin{figure}
\begin{center}
\includegraphics{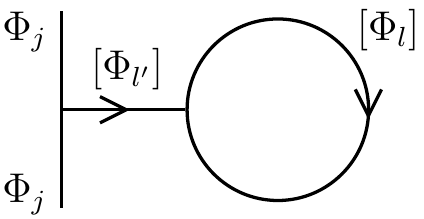}
\caption{Current block for the torus two-point function  in the $z\rightarrow e^{2\pi i} z$ eigenstate basis.}
\label{twopointfig}
\end{center}
\end{figure}
The current blocks of the two-point function $\langle
\Phi_j(z)\Phi_j(0)\rangle$ of a primary field $\Phi_j$ on the torus
are shown in figure~\ref{twopointfig}. The field running in the torus
is denoted by $[\Phi_{l}]$, while the field $[\Phi_{l'}]$ connects the
propagator to the loop.\footnote{Square brackets $[...]$ denote all
  fields allowed by the fusion rules, {\em i.e.}\ the primary and all
  its current algebra descendants.}  For a given field $ \Phi_{j}$,
the connecting fields allowed by the fusion rules are $[\Phi_{l'}]$
with $0 \leq l' \leq \min(2j,k-2j)$.\footnote{The $SU(2)$ fusions rules are
  given in appendix~\ref{appB}.} To each field $[\Phi_{l'}]$, the
fusion rules allow $k+1-2l'$ fields $[\Phi_{l}]$
($l=l'/2,...,(k-l')/2$) in the loop.  Each two-point correlator
therefore contains
\begin{align}
n=\sum_{l'=0}^{\min(2j,k-2j)} (k+1-2l') = (2j+1)(k-2j+1) 
\end{align}
current blocks.

A simple class of two-point functions is $\langle
\Phi_j(z)\Phi_j(0)\rangle$ with spin $j=\frac{k}{2}$ (corresponding to
$h=\frac{k}{4}$), which is the maximal spin at level $k$. In this case
the only field in the intermediate channel allowed by the fusion rules
is the identity $(l'=0)$, and there are $k+1$ fields $[\Phi_l]$
($l=0,\frac{1}{2},...,\frac{k}{2}$) in the loop. As argued in the
introduction, this class of two-point functions is constructed from
the characters of the $SU(2)$ WZW model and the torus prime form.  The
level-$k$ characters of the $SU(2)$ WZW model on the torus were
derived in \cite{KP} and are given by
\begin{align}\label{char}
\chi_{l}^{(k)} (\tau) &\equiv \lim_{z \rightarrow 0}  {\chi^{(k)}_l(z,\tau)} = \lim_{z \rightarrow 0} 
\frac{\Theta_{2l+1, k+2}(z, \tau) 
- \Theta_{-2l-1, k+2}(z, \tau)}  
{\Theta_{1, 2}(z, \tau)-\Theta_{-1, 2}(z, \tau)} \,,\qquad 0 \leq l \leq \textstyle\frac{k}{2}\,,
\end{align}
where the Kac-Peterson theta functions $\Theta_{m, k}(z, \tau)$ are
defined as
\begin{align}
\Theta_{m, k}(z, \tau) \equiv  \sum_{n\in \ZZ+\frac{m}{2k}} 
q^{k n^2} x^{k n}  \,,
\end{align}
and $q=e^{2\pi i \tau}$, $x=e^{2\pi i z}$. The characters
$\chi_{l}^{(k)}$ formally correspond to the torus one-point function
of the identity $\langle \Phi_0 \rangle_l$ with a primary field
$\Phi_l$ running in the loop.

The prime form on the torus $E(z,\tau)$ is defined by
\begin{align}
E(z,\tau)=\frac{\theta_1(z,\tau)}{\theta'_1(0,\tau)}\,,
\end{align}
where $\theta_1(z,\tau)$ is the first Jacobi theta function.

We summarize our claim for this class of two-point functions in the
following theorem:
\begin{thm}\label{th1}
  The current blocks of the level-$k$
  spin-$\frac{k}{2}$-spin-$\frac{k}{2}$ two-point function on the
  torus $(k \in \mathbb{N})$,
\begin{align}
\langle &\Phi_{\frac{k}{2}}(z,\bar z)
\Phi_{\frac{k}{2}}(0, 0) \rangle
=  \sum^{k/2}_{ l=0,\, l \in \mathbb{N}_0/2}
\left\vert f_l(z,\tau) \right\vert^2 \,,
\end{align}
are 
\begin{align}
f_l(z,\tau)= \frac{\chi^{(k)}_l(z,\tau)}{E(z,\tau)^{2h}} \,,\qquad \textstyle l=0,\frac{1}{2},...,\frac{k}{2}\,, \label{theorem}
\end{align}
where $\chi^{(k)}_l(z,\tau)$ is given by (\ref{char}), and
$h=\frac{k}{4}$ is the conformal dimension of the
primary~$\Phi_\frac{k}{2}$.
\end{thm}

We remark here that the theorem is formulated for the $A$-type modular invariant  theory, which exists
for all $k\geq 1$.  In this case the two-point function is simply the sum over the product of holomorphic and
anti-holomorphic current blocks. For $k=4\rho$ ($\rho \in \mathbb{N}$), 
there is also a $D$-type modular invariant theory, and the torus two-point function is 
\begin{align}
\langle &\Phi_{\frac{k}{2}}(z,\bar z) \Phi_{\frac{k}{2}}(0, 0) \rangle
=  \sum^{(k-2)/4}_{ l=0,\, l \in \mathbb{N}_0/2} \left\vert f_l(z,\tau) + 
 f_{\frac{k}{2}-l}(z,\tau) \right\vert^2 +2  \left\vert f_{\frac{k}{4}}(z,\tau) \right\vert^2  \,,
\end{align}
where $\Phi_{\frac{k}{2}} \equiv \Phi_{\frac{k}{2}, \frac{k}{2}}$. Likewise, we  construct the 
corresponding two-point functions for the other $D$- and $E$-type modular invariant 
theories following Table~2 in \cite{Cappelli}.

\setcounter{equation}{0}
\section{Factorization limits and monodromy properties}\label{sec3}

Before we will give a proof of theorem~\ref{th1} in the next section,
we will first examine the factorization limits and monodromy
properties of the current blocks (\ref{theorem}).

\subsection{Factorization in the limits $z \rightarrow 0$ and $q\rightarrow 0$}

In the limit $z\rightarrow 0$, the prime form $E(z,\tau)\rightarrow z$
and the current blocks (\ref{theorem}) trivially factorize as
\begin{align}
  f_l(z, \tau) \stackrel{z\rightarrow 0}{\longrightarrow}
  \frac{1}{z^{2h}} \cdot {\chi^{(k)}_l(\tau)} \,,
\end{align} 
{\em i.e.}\ into the two-point function on the sphere
\cite{FZ} and the character ${\chi^{(k)}_l(\tau)}$ \cite{KP}, as
expected by pinching the intermediate channel in
figure~\ref{twopointfig}.

\medskip 
\begin{figure}
\begin{center}
\includegraphics{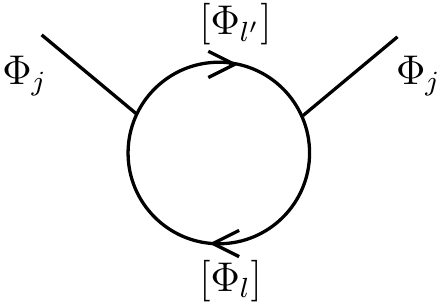}
\caption{Current block for the torus two-point function  in the $z\rightarrow z+1$ eigenstate basis.}
\label{twopointfig2}
\end{center}
\end{figure}
Another interesting limit is $q \rightarrow 0$ ({\em i.e.}\
$\tau \rightarrow i\infty$), which corresponds to pinching the torus
along the a-cycle. In this limit the appropriate representation of the
current blocks $f_l$ is given in figure~\ref{twopointfig2}.  Here the $\Phi_j$'s are the external states, and 
$[\Phi_l]$ and $[\Phi_{l'}]$ denote the primaries (and their descendants) running between the two
insertions. As argued
in \cite{Brustein, Kiritsis}, in this limit a general current block
$f_l$ is expected to degenerate into a four-point function on the
sphere with two $\Phi_{l'}$'s and two $\Phi_j$'s as external states
and $\Phi_l$ in the intermediate channel,
\begin{align}
f_l(z,\tau) \stackrel{q\rightarrow 0}{\longrightarrow}   q^{h_l-\frac{c}{24} }  x^h \langle l' | \Phi_j(1) \Phi_j(x) |l'\rangle_l \,.
\end{align}
Here $x=e^{2\pi i z}$ is the cross-ratio and $c$ the central charge of
$SU(2)$ at level~$k$.  Writing the four-point function as a product of
three-point vertices and expanding around $x=0$, we get the following
behavior:
\begin{align}
f_l  \stackrel{q\rightarrow 0}{\longrightarrow}   q^{h_l-c/24} x^{h_{l'}-h_l} \left(1+ K_{1,2l+1} x + K_{2,2l+1} x^2 + ...\right) \,, \label{flx}
\end{align}
where $h_{l}$ and $h_{l'}$ are the conformal dimensions of $\Phi_l$
and $\Phi_{l'}$, respectively.  The coefficients of the subleading
terms $K_{i,2l+1}$ ($i=1,2,...$) can in principle be determined by
elementary methods \cite{Durganandini}.

Let us now consider the $q\rightarrow 0$ limit of the current blocks
(\ref{theorem}).  In the double limit $q \rightarrow 0$, $x
\rightarrow 0$, the prime form and functions $\chi^{(k)}_l(z,\tau)$
given by (\ref{char}) behave to leading order in $q$ as
\begin{align}
E(z,\tau) &\sim x^{-1/2} (1-x+...)
 \,,\\
\chi^{(k)}_{l}(z,\tau) &\sim q^{h_l-c/24} x^{-l} (1+ {\cal O}(x) ) 
 \,,
\end{align}
such that the current blocks (\ref{theorem}) behave as 
\begin{align}
  f_l(z,\tau)  \sim q^{h_l-c/24} x^{h - l} \left(1+ \tilde K_{1,2l+1} x 
  + \tilde K_{2,2l+1} x^2 + ...\right) \,.\label{expansion}
\end{align}
To leading order in $x$, we get agreement with (\ref{flx}), if
\begin{align}
h-l = h_{l'}-h_l \,.\label{cond} 
\end{align}
A general current block does not satisfy this condition. However,
(\ref{cond}) is satisfied for the class of two-point functions of
primary operators $\Phi_j(z)$ with $j=\frac{k}{2}$.  In that case, the
conformal dimension is $h=\frac{k}{4}$ and $l+l'=j$ such that both
sides of (\ref{cond}) become equal to $\frac{k}{4}-l$, and we find
that to leading order in $x$ (\ref{theorem}) has the correct
asymptotics in the limit $q\rightarrow 0$. In appendix~\ref{appDE} we
also find agreement to next-to-leading order in $x$ by showing $K_1 =
\tilde K_1$ for a few special cases.

\subsection{Monodromy properties}

The complete two-point function is a uniform, doubly periodic function
on the torus. Its current blocks however have non-trivial monodromy
properties, {\em i.e.}\ there exist mono\-dromy matrices $M_\sigma$
and $M_\tau$ such that under translations $z\rightarrow z+1$ and
$z\rightarrow z+\tau$, $f_l \rightarrow (M_\sigma)_{lm} f_m$ and $f_l
\rightarrow (M_\tau)_{lm} f_m$, respectively \cite{MMS2}.  As shown in
\cite{MMS}, there exists a basis of functions $f_l$ in which
$M_\sigma$ is diagonal with eigenvalues $e^{2\pi i (h_l-h_{l'}) }$
such that
\begin{align}
f_l(z+1,\tau)= e^{2\pi i (h_l-h_{l'}) }f_l(z,\tau)\,.\label{expectedmono}
\end{align}
This basis corresponds to choosing the holomorphic blocks as in
figure~\ref{twopointfig2}. Here $h_l$ and $h_{l'}$ are the conformal weights
of the primaries $\Phi_l$ and $\Phi_{l'}$ running between the two
insertions.  By modular invariance, the eigenvalues of $M_\tau$ are
the same as those of $M_\sigma$, but $M_\sigma$ and $M_\tau$ cannot be
diagonalized simultaneously \cite{MMS2}.

Let us consider the monodromy properties of the current blocks
(\ref{theorem}). The behavior under monodromy transformations can be
deduced from that of the Kac-Peterson theta functions given by
(\ref{monodromy}) in appendix~\ref{appA}.  Under the monodromy $z
\rightarrow z+1$, the prime form and Kac-Peterson characters pick up
some phases,
\begin{align}
E(z+1,\tau) &= -E(z,\tau)\,,\nonumber\\
\chi_l(z+1,\tau) &=(-1)^{2l} \chi_l(z,\tau) \,,
\end{align}
such that the current blocks (\ref{theorem}) transform as
\begin{align}
f_l(z+1,\tau)= (e^{\pi i})^{2(l-h)} f_l(z,\tau)\,. \label{shift1}
\end{align}
Comparing this with (\ref{expectedmono}), we find again the
requirement (\ref{cond}), which is satisfied for $j=\frac{k}{2}$, as
argued above.

\medskip

Under the monodromy $z\rightarrow z+\tau$, the prime form and
Kac-Peterson characters transform as
\begin{align}
E(z+\tau,\tau) &=
 e^{-\pi i} \frac{1}{x q^\frac{1}{2}} E(z,\tau)\,,\nonumber\\ 
\chi_l(z+\tau,\tau) &=
x^{-\frac{k}{2}} q^{-\frac{k}{4}}  \chi_{\frac{k}{2}-l}(z,\tau) \,,
\end{align}
and therefore
\begin{align}
f_l(z+\tau,\tau)=
(e^{\pi i})^{2h} x^{2h -\frac{k}{2} } q^{h-\frac{k}{4}}   f_{\frac{k}{2}-l}(z,\tau)\,.
\end{align}
For the case $j=\frac{k}{2}$ $(h=\frac{k}{4})$, the transformation
becomes independent of $z$ and $\tau$,
\begin{align}\label{shift2}
f_l(z+\tau,\tau)=
(e^{\pi i})^{2h}   f_{\frac{k}{2}-l}(z,\tau)\,,
\end{align}
a fact which will be of importance in the proof.
\medskip

In summary, $M_\sigma$ and $M_\tau$ are $k+1$-dimensional matrices of
the type
\begin{align}
M_\sigma =
\begin{pmatrix}
(e^{\pi i})^{-2h} &  & \\
&\ddots &\\
& &(e^{\pi i})^{\frac{k}{2}-2h}
\end{pmatrix}
\,,\qquad
M_\tau =(e^{\pi i})^{2h}
\begin{pmatrix}
 &  & 1\\
&\reflectbox{$\ddots$} &\\
1& &
\end{pmatrix} \,,
\end{align}
with $h=\frac{k}{4}$. For instance, for $k=1$, $M_\sigma=i
\sigma_3={\rm diag}(e^{i \pi /2}, e^{-i \pi/2})$ and
$M_\tau=i\sigma_1$ in agreement with \cite{MMS}. It can be shown that,
for $h=\frac{k}{4}$, $M_\tau$ has the same eigenvalues as $M_\sigma$,
as required.  The current blocks (\ref{theorem}) thus have the proper
monodromy properties with respect to the transformations $z\rightarrow
z+1$ and $z\rightarrow z+\tau$.

\setcounter{equation}{0}
\section{Differential equations for the two-point conformal blocks on the torus}\label{sec4}

In this section we will prove theorem~\ref{th1} by showing that the
current blocks (\ref{theorem}) \mbox{satisfy} 
an appropriate  Knizhnik-Zamolodchikov-like differential equation. 
Such  differential equations were first constructed in \cite{MMS} using
a combination of the current Ward identity and 
a current algebra null-vector relation. 
It was then noticed \cite{MMS2} (see also \cite{Kiritsis}) 
that the same equations can be derived exclusively from modular invariance
and the Wronskians of the current blocks.

\subsection{Brief review of the method}

Let us consider the two-point correlator of a primary field $\Phi_j$
at level $k$. It is constructed from $n$ holomorphic conformal blocks
$f_l$ $(l=1,...,n)$, which are the solutions of a single $n$th-order
differential equation of the form
\begin{align}\label{diff}
W(z,q)\partial^{n}f_l+\underset{k=0}{\overset{n-1}{\sum}}(-1)^{n-k}W_{k}(z,q)\partial^{k}f_l=0\,,
\end{align}
where $\partial\equiv\partial_z$, $W=W_{n}$ and $W_{k}$ is defined by
\begin{align}
W_{k}=\det\left[\begin{array}{ccccc}
f_{1} &  & \cdots &  & f_{n}\\
\partial_{z}f_{1} &  & \cdots &  & \partial_{z}f_{n}\\
\vdots &  &  &  & \vdots\\
\partial_{z}^{k-1}f_{1} &  & \cdots &  & \partial_{z}^{k-1}f_{n}\\
\partial_{z}^{k+1}f_{1} &  & \cdots &  & \partial_{z}^{k+1}f_{n}\\
\vdots &  &  &  & \vdots\\
\partial_{z}^{n}f_{1} &  & \cdots &  & \partial_{z}^{n}f_{n}\end{array}\right]\,,
\end{align}
see \cite{MMS2}  for the derivation of (\ref{diff}).

The Wronskians $W_k$ are meromorphic functions of $z$ with a single
pole at $z=0$ and invariant under monodromy transformations
\cite{MMS2}.  Since the derivatives of the Weierstrass $\wp$-function
form a basis for meromorphic functions on the torus with poles at $z =
0$, the $W_k$'s can be expanded as
\begin{align}
W_k = 
 \sum_{r=0}^{p+n-k-2}
 A_r^{(k)}(\tau)  \partial^{p+n-k-2-r} \wp(z,\tau) +   A_{p+n-k}^{(k)}(\tau) \,,
\end{align}
where $p$ is the order of the pole of $W=W_n$ at $z=0$.\footnote{The
  behavior $W \sim z^{-p}$ near $z=0$ can be determined from the
  asymptotic behavior of the current blocks near $z=0$, see
  (\ref{asympbeh}) below.}  $A_{r}^{(k)}(\tau)$ are modular forms of
weight~$r$. Note that $A_{r}^{(k)}(\tau)=0$ for $r=2$ and odd $r$,
since there is no modular form with weight~2 or odd weight.  In the
simplest case, when the space $M_r$ of modular forms of weight $r$ is
one-dimensional ({\em e.g.}\ for $r=4,6,8,10,14$), we may express
$A_{r}^{(k)}(\tau)$ in terms of the Eisenstein series,
$A_{r}^{(k)}(\tau)= a^{(k)}_r E_r$ with $a^{(k)}_r$ some
$\mathbb{C}$-valued coefficient.

The constants $a^{(k)}_r$ can be found by studying the differential
equation (\ref{diff}) both at $z=0$ and $z=i \infty$.  In particular,
the coefficients $a_{0}^{(k)}$ are obtained by analyzing (\ref{diff})
around $z=0$.  Near $z=0$, the Weierstrass-$\wp$ function is
$\wp(z,\tau) \approx z^{-2}$.
Substituting $f(z) \sim z^\mu$ into (\ref{diff}), we get the linear equation
\begin{align}
  \sum_{k=0}^n (-1)^{n-k} a_{0}^{(k)} \left(\prod_{j=0}^{k-1} (\mu-j) \right) 
  \left(\prod_{l=0}^{p+n-k-3} (-2-l) \right)=0\,,\label{lineareqn}
\end{align}
where the second product equals one if $p+n-k-3<0$.

Let us now determine the behavior of the current blocks $f_l \sim
z^\mu$ near $z=0$. For this, we define $n_{l'}=k+1-2l'$ as the number
of conformal blocks $f_l$ with the field $[\Phi_{l'}]$ in the
intermediate channel, see figure~\ref{twopointfig}.  Their asymptotic
behavior near $z=0$ is
\begin{align}
z^{-(2h-h_{l'} - l')}, z^{-(2h-h_{l'} - l'-2)}, ..., z^{-(2h-h_{l'} - 2(n_{l'}-1))}\,.\label{asympbeh}
\end{align}
For $l'=0$, the leading behavior is just $z^{-2h}$. For $l'>0$, the
one-point function of $\Phi_{l'}$ vanishes on the torus. In fact, all
descendants of $\Phi_{l'}$ up to level $l'-1$ have vanishing one-point
functions~\cite{Kiritsis}. The first non-vanishing one-point function
corresponds to the state $J_{-1}^{a_1} \cdots
J_{-1}^{a_{l'}}|\Phi_{l'}\rangle$, which has conformal dimension
$h_{l'}+l'$. For $W_k$ to be non-vanishing, we also keep the
non-leading terms of the current blocks, corresponding to the
descendants $J_{-1}^{a_1} \cdots J_{-1}^{a_{l'+2}}|\Phi_{l'}\rangle$,
$J_{-1}^{a_1} \cdots J_{-1}^{a_{l'+4}} |\Phi_{l'}\rangle$ etc.  Since
$n=\sum_{l'} n_{l'}$, we find $n$ values for $\mu$ and therefore $n$
independent linear equations of the type (\ref{lineareqn}), enough to
determine the $n$ coefficients $a_0^{(0)}, ..., a_0^{(n-1)}$ (Since
the differential equation (\ref{diff}) can be multiplied by a
constant, we may set the coefficient $a^{(n)}_0=1$).  The remaining
coefficients $a_{r}^{(k)}$ ($r \neq 0$) are obtained in a similar way
by expanding around $z=i \infty$.

\medskip

In appendix \ref{appDE}, we list the differential equations
(\ref{diff}) for the current blocks of the torus two-point function
$\langle \Phi_\frac{k}{2}(z)\Phi_\frac{k}{2}(0)\rangle$ at level $k$
for $k=1,2,...,5$.  As a first test of our solution, we substitute
(\ref{theorem}) (for all $k\leq 5$ and all
$l=0,\frac{1}{2},...,\frac{k}{2}$) into (\ref{diff}) and expand the
left-hand side around $q=0, z=0$.  We find that (\ref{theorem})
satisfies (\ref{diff}) up to the order we expanded in $z$ and
$q$.\footnote{Using computer algebra (mathematica), this can be done
  to arbitrary orders in $z$ and $q$ and is only limited by computer
  power.}  We now give a more rigorous proof of theorem~\ref{th1}.

\subsection{Proof of  theorem \ref{th1}}

For the proof we will need a variation of Liouville's theorem for
elliptic functions:
\begin{thm} 
  An entire elliptic (i.e.\ doubly periodic) function is constant.
\end{thm}
\medskip

For a  special class of quasi-periodic
functions, this theorem can be extended to the
\begin{prop}\label{lemma}
{\em Let $g(z)$ be an entire quasi-periodic function $g(z)$,
\begin{align}
g(z+ 1) = e^{i\phi_{1}} g(z) \,,\qquad
g(z+ \tau) = e^{i\phi_{2}} g(z)\,,\label{quasi}
\end{align}
whose multipliers $e^{i\phi_{i}}$ $(i=1,2)$ are just phases ($\phi_{i}=const.$).
If  $g(0)=0$, then $g(z)\equiv 0$ for all $z$.\footnote{We thank Yuri Aisaka for discussions which led to the formulation of this proposition.} }
\end{prop}

Proof of proposition~\ref{lemma}: In this case the absolute values of
$g(z)$ are invariant under shifts $z\rightarrow z+1$ and $z\rightarrow
z+\tau$, {\em i.e.}\ $|g(z)|=|g(z+1)|=|g(z+\tau)|$ as for elliptic
functions.  Therefore, if $g(z)$ is holomorphic and its absolute value
is bounded inside the fundamental parallelogram, by the
quasi-periodicity of $g(z)$, it is bounded for all $z$ and thus
constant. The requirement $g(0)=0$ sets the constant to zero.
$\quad\square$ 
\medskip

To prove theorem \ref{th1}, we substitute the $n=k+1$ current blocks
of the spin-$\frac{k}{2}$-spin-$\frac{k}{2}$ two-point function,
\begin{align}
f_l(z,\tau)= \frac{\chi^{(k)}_l(z,\tau)}{E(z,\tau)^{2h}} \,,\qquad \textstyle l=0,\frac{1}{2},...,\frac{k}{2}\,, 
\qquad h=\frac{k}{4}\,,
\end{align}
with $\chi^{(k)}_l(z,\tau)$ as in (\ref{char}), into the differential
equation (\ref{diff}) for a fixed but arbitrary level~$k$. We then
define the function $g_l=g_l(z,\tau)$ as the left-hand side of
(\ref{diff}),
\begin{align}
g_l \equiv {\partial^{n} f_l}+\underset{r=0}{\overset{n-1}{\sum}}(-1)^{n-r}\frac{W_{r}}{W} {\partial^{r}f_l}\,,
\end{align}
where the Wronskians are those of the
spin-$\frac{k}{2}$-spin-$\frac{k}{2}$ two-point function at level~$k$.
The Wronskians are in principle known for all levels $k$ and can be
constructed as described in the previous subsection. For the proof we
do not need to do this explicitly. It will turn out to be enough to
know a few properties of the Wronskians. We need to show that
$g_l(z,\tau) \equiv 0$ for all $z$ and $\tau$.  \medskip

It follows from proposition~\ref{lemma} that $g_l(z,\tau) \equiv 0$
for all $z$ and $\tau$ if $g_l(z,\tau)$
\begin{itemize}
\item[(i)] is a quasi-periodic function of the type (\ref{quasi}) with
  periods $1$ and $2\tau$,
\item[(ii)] has no poles in the $z$-plane inside the parallelogram
  with sides $1$ and $2\tau$,
\item[(iii)] vanishes at $z = 0$.
\end{itemize}

\subsubsection*{(i) Quasi-periodicity of \boldmath{$g_l$}:}
The function $g_l$ inherits the monodromies of $f_l$,
\begin{align}
g_l(z+1,\tau)= e^{2\pi i (h_l-h_{l'}) }g_l(z,\tau)\,,\qquad
g_l(z+\tau,\tau)= (e^{\pi i })^{2h} g_{\frac{k}{2}-l}(z,\tau)\,,
\end{align}
as follows from the following observations:
\begin{itemize}
\item The derivatives of the holomorphic blocks $f_l$ have the same
  monodromy properties as $f_l$. This follows from the fact that the
  monodromies of $f_l$ are independent of~$z$, see (\ref{shift1}) and
  (\ref{shift2}) for $h=\frac{k}{4}$.  The derivatives ${\partial^r
    f_l}$ are therefore quasi-periodic with the same monodromies as
  $f_l$.
  
\item The Wronskians $W$ and $W_r$ are monodromy invariant \cite{MMS2}. 
\end{itemize}
Therefore, $g_l(z+2\tau,\tau)= (e^{\pi i })^{4h} g_{l}(z,\tau)$, {\em i.e.}\ the 
periods of $g_l$ are $1$ and $2\tau$.

\subsubsection*{(ii) Poles of  \boldmath{$g_l$}:}

The function $g_l$ can have poles at most at $z=m+n\tau$ ($m, n \in
\mathbb{Z}$). This follows from the fact that the individual terms in
$g_l$, which are of the type $\frac{W_{r}}{W} {\partial^{r}f_l}$, have
poles at $z=m+n \tau$:
\begin{itemize}
\item The only poles of $f_l(z,\tau)$ are at $z=m+n \tau$. Note that
  the theta functions are holomorphic in $z$ and the denominator of
  $\chi_l^{(k)}(z,\tau)$,
\begin{align}
  \Theta_{1, 2}(z, \tau)-\Theta_{-1, 2}(z, \tau) = i
  \theta_1(z,\tau)\,,
\end{align}
is zero at $z=m+n \tau$. A similar argument holds for the inverse of
the torus prime form $E(z,\tau)$.
\item  The derivatives ${\partial^r f_l}$ have poles at the same locations as
the functions $f_l$.
\item In general, $W_r/W$ is a meromorphic single-valued function of
  $z$ with poles at $z = 0$ and at the locations of the zeros of $W$
  \cite{MMS2}.
  However, in the differential equations for
  spin-$\frac{k}{2}$-spin-$\frac{k}{2}$ two-point functions, $W$~is a
  (non-vanishing) constant \cite{MMS2}. This leaves only $z=0$ as a
  possible location for a pole of $W_r/W$.
\end{itemize}
Even though the individual terms have poles at $z=m+n \tau$, the total
sum of the pole contributions vanishes in $g_l$, as follows from
property (iii).
 
\subsubsection*{(iii)  \boldmath{$g_l=0$} at \boldmath{$z=0$}:}
\begin{itemize}
\item Near $z=0$, the current blocks $f_l$ behave like $z^{-k/2}$
  ($=z^{-2h}$). The solutions $f_l$ therefore trivially satisfy
  (\ref{diff}) near $z=0$, since the coefficients $a_0^{(k)}$ are
  constructed such that $\mu=-2h$ is a solution of (\ref{lineareqn}).
  In other words, (\ref{diff}) is constructed such that $f \sim
  z^{-2h}$ is a solution of it near $z=0$. Thus, $g_l=0$ at $z=0$
  (This excludes poles of $g_l$ at $z=m+n \tau$ ($m, n \in
  \mathbb{Z}$) such that $g_l$ does not have any poles).
\end{itemize}

\noindent
In conclusion, by (i), (ii) and (iii), $g_l$ is a quasi-periodic
function of the type (\ref{quasi}), free of poles and $g_l(0)=0$. By
proposition~\ref{lemma}, $g_l \equiv 0$ for all~$z$.

\medskip 
A final remark on the overall normalization of the current
blocks as a function of $\tau$ is in order.  The differential equation
(\ref{diff}) determines the current blocks only up to an overall
$\tau$-dependent factor and cannot be used to check the normalization
of $f_l$. However, the current blocks $f_l$ have the correct
$\tau$-depen\-dence by construction.  This follows from the fact that
in the limit $z\rightarrow 0$ the total torus
spin-$\frac{k}{2}$-spin-$\frac{k}{2}$ two-point function factorizes
into the corresponding two-point function on the sphere and the
partition function $Z=\sum_l |\chi_l|^2$.  The transformation
properties of the torus two-point function under the modular
transformations $S: \tau \rightarrow -\frac{1}{\tau}$ and $T: \tau
\rightarrow \tau + 1$ therefore follow directly from those of the
characters.  $\quad\square$

\setcounter{equation}{0}
\section{Conclusions}\label{sec5}

The main result of this paper is the expression (\ref{theorem}) for
the current blocks of the $SU(2)$
spin-$\frac{k}{2}$-spin-$\frac{k}{2}$ torus two-point function at
arbitrary level $k$. The theorem was proven using the differential
equation technique developed in \cite{MMS, MMS2}.

One may wonder whether the expression (\ref{theorem}) also describes
the current blocks of other torus $SU(2)$ two-point functions which
have the identity $[\Phi_0]$ in the intermediate channel. Consider for
instance the current block $f_{l=1}$ of the
spin-$\frac{1}{2}$-spin-$\frac{1}{2}$ two-point function at
level~$k=2$.  This block has $[\Phi_0]$ in the intermediate channel
and $[\Phi_1]$ in the loop.  For this case, the expression
(\ref{theorem}) with $\chi_l(z,\tau)$ given by (\ref{char}) has the
wrong factorization property in the limit $q \rightarrow 0$, even
though its factorization in the limit $z\rightarrow 0$ is correct.
Note that the important condition (\ref{cond}) is not satisfied since
$h-l= \frac{3}{16}-1=-\frac{13}{16} \notin \{\pm\frac{3}{16}, \pm
\frac{5}{16}\}$, which is the set of values $h_{l'}-h_l$ for the four
conformal blocks of this correlator. It would therefore be interesting
to study how the character $\chi_l(\tau)$ is properly extended to a
function $\widetilde \chi_l(z,\tau)\neq \chi_l(z,\tau)$ for such
current blocks. Eventually it would be interesting to find an
expression for the current blocks of all $SU(2)$ torus two-point
functions and higher $n$-point functions.

\section*{Acknowledgments}

We thank Yuri Aisaka, Matthias Gaberdiel, Terry Gannon, Volker Schomerus
and Cosimo Restuccia for helpful comments and discussions related to this work.
 I.K.\ is grateful to Samir Mathur for email correspondence. P.K.\ thanks the 
Deutsches Elektronen-Synchrotron (DESY) for its hospitality
during the summer student programme 2012 and Frank Tackmann for initial support.
\bigskip

\section*{Appendix}

\appendix
\setcounter{equation}{0}
\section{Theta functions} \label{appA}
\subsection{Theta functions with characteristics} 

The (first-order) theta function with characteristics is defined as
\cite{Fay}
\begin{align}
  \thetaR{\delta}{\varepsilon}(z,\tau) = \sum_{n \in \ZZ} e^{2\pi i
    (\frac{1}{2} (n+\delta)^2 \tau + (z+\varepsilon)(n+\delta))}\,.
\end{align}
Under translations $z\rightarrow z+a+b\tau$, it transforms as
\begin{align}
  \thetaR{\delta}{\varepsilon}(z+a+b\tau,\tau) = e^{2\pi i
    (-\frac{1}{2} b^2 \tau - b z + a\delta - b \varepsilon)}
  \thetaR{\delta+b}{\varepsilon}(z,\tau) \label{idR}\,.
\end{align}

\subsection{Kac-Peterson theta functions for $SU(2)$}

The Kac-Peterson theta functions are
\begin{align}
  \Theta_{m,k}(z,\tau) \equiv \thetaR{\frac{m}{2k}}{0}(k z, 2 k \tau)
  = \sum_{n\in \ZZ+\frac{m}{2k}} q^{k n^2} x^{k n} \,,
\end{align}
where $q=e^{2\pi i \tau}$, $x=e^{2\pi i z}$. 
Using (\ref{idR}), we find the monodromy transformations
\begin{align}\label{monodromy}
\Theta_{m,k}(z+1, \tau) &= e^{\pi i  m } \Theta_{m,k}(z, \tau)\,, \nonumber\\
\Theta_{m,k}(z+\tau, \tau) &= e^{\pi i (-   k \tau/2 -   k z )} \Theta_{m+k,k}(z, \tau)\,.
\end{align}
Under modular transformations, they transform as
\begin{align}
T:\quad&\Theta_{m,k}(z,\tau + 1) = e^{2\pi i \frac{m^2}{4k}} \Theta_{m,k}(z, \tau)\,,\label{T}\\
S:\quad&\Theta_{m,k}(z/\tau, -1/\tau) =  e^{2\pi i (z^2/4\tau)} (-i \tau)^{1/2} 
\sum_{m'=-k+1}^k  B^{(k)}_{mm'} \Theta_{m',k}(z, \tau)  \,, \label{S}\\
 &\qquad\qquad\,\,\,\, B^{(k)}_{mm'} = \frac{1}{\sqrt{2k}} e^{\pi\,imm'/k} \,.\nonumber
\end{align}
Moreover,
\begin{align}
\quad&\Theta_{m,k}(-z, \tau) = \Theta_{-m,k}(z, \tau) \,,\qquad
\Theta_{m+2k,k}(z, \tau) = \Theta_{m,k}(z, \tau) \,.
\end{align}
The prime form $E(\tau, z)$ transforms as
\begin{align}
E(z, \tau+1) &= E(z, \tau)\,,\\
E( z/\tau, -1/\tau) &= \frac{e^{\pi i (z^2/2\tau)}}{\tau} E(z, \tau) \,.
\end{align}

\setcounter{equation}{0}
\section{Fusion rules}\label{appB}

The fusion rules for the $SU(2)$ WZW model at level $k$ can be
expressed in terms of the fusion matrices 
$N^{(j_3)}_{j_1j_2}=N^{j_3}_{j_1j_2}$ as
\begin{align} [\phi_{j_1}] \times [\phi_{j_2}] = \sum_{j_3}
  N^{j_3}_{j_1j_2} [\phi_{j_3}] \,,
\end{align}
where $N^{j_3}_{j_1j_2} = 1$ if $|j_1-j_2| \leq j_3 \leq {\rm
  min}(j_1+j_2, k-j_1-j_2)$, otherwise $N^{j_3}_{j_1j_2} = 0$ \cite{FZ}.

\setcounter{equation}{0}
\section{Differential equations }\label{appDE}

In this appendix we list the differential equations for the current
blocks of the torus two-point function
$\langle\phi_{{k}/{2}}(z)\phi_{{k}/{2}}(0)\rangle$ at level $k$ for
$k=1, 2,...,5$.  We also give the coefficients $K_{1,2l+1}$ in the
asymptotic expansion of the current blocks,
\begin{align}
f_l(z,\tau)  \sim q^{h_l-c/24} x^{h_{l'}-h_l} \left(1+ K_{1,2l+1} x + ...\right) \,. 
\end{align}
The coefficients $K_{1,1}=\frac{k}{2}$ and $K_{1,2}=\frac{k}{2}+1$
follow from our procedure and agree with the corresponding
coefficients $\tilde K_{1,1}$ and $\tilde K_{1,2}$ of the small $q$
and $x$ expansion of the solution (\ref{theorem}), as given by
(\ref{expansion}).  The remaining coefficients can be computed by
elementary methods \cite{Durganandini}.

\medskip

The differential equations for the spin-$\frac{k}{2}$-spin-$\frac{k}{2}$ current blocks are:\\

$k=1, j=1/2$:
\begin{align}
& f''(z,\tau)-\frac{3}{4} \wp(z,\tau) f(z,\tau)=0\label{eq:k=00003D1,j=00003D1/2}\,,\\
& \textstyle K_{1,1}=\frac{1}{2}\,,\quad K_{1,2}=\frac{3}{2} \,,
\end{align}

$k=2, j=1$:
\begin{align}
& f^{(3)}(z,\tau)-3\wp(z,\tau) f'(z,\tau)-\frac{3}{2}\wp'(z,\tau)f(z,\tau) =0\label{eq:k=00003D2,j=00003D1}\,,\\
&\textstyle K_{1,1}=1\,,\quad K_{1,2}={2} \,,\quad K_{1,3}={2} \,,
\end{align}

$k=3, j=3/2$:
\begin{align}
&f^{(4)}(z,\tau)-\frac{15}{2}\wp(z,\tau) f''(z,\tau)-\frac{15}{2} \wp'(z,\tau) f'(z,\tau)  \nonumber\\
&\quad+ \left( \frac{9\pi^{4}}{16}\text{E}_{4}(\tau)- \frac{45}{32} \wp''(z,\tau) \right) f(z,\tau)=0\label{eq:k=00003D3}\,,\\
&\textstyle K_{1,1}=\frac{3}{2}\,,\quad K_{1,2}=\frac{5}{2} \,,\quad K_{1,3}=\frac{20}{8}  \,,\quad K_{1,4}=\frac{20}{8}\,,
\end{align}

$k=4, j=2$:
\begin{align}
& f^{(5)}(z,\tau)-15\wp(z,\tau)f^{(3)}(z,\tau)-\frac{45}{2} \wp'(z,\tau) f''(z,\tau) 
\nonumber\\
&\quad+\left(4\pi^{4}\text{E}_{4}(\tau)- \frac{15}{2} \wp''(z,\tau)\right) f'(z,\tau) =0\,,\label{eq:k=00003D4}\\
&\textstyle K_{1,1}={2}\,,\quad K_{1,2}=3 \,,\quad K_{1,3}=3  \,,\quad K_{1,4}=3 \,.\quad K_{1,5}=3 \,,
\end{align}

$k=5, j=5/2$:
\begin{align}
 &f^{(6)}(z,\tau)-\frac{105}{4}  \wp(z,\tau ) f^{(4)}(z,\tau) -\frac{105}{2} \wp'(z,\tau) f^{(3)}(z,\tau)\nonumber\\
 &\quad+  \left( \frac{259 \pi ^4}{16}\text{E}_4(\tau) -\frac{735}{32} \wp''(z,\tau )\right) f''(z,\tau)+\frac{105}{32} \wp^{(3)}(z,\tau )f'(z,\tau) \nonumber\\
 &\quad+\left( -\frac{735 \pi ^4}{64} \text{E}_4(\tau) \wp(z,\tau )-\frac{5 \pi ^6}{16}\text{E}_6(\tau)
  + \frac{1155}{512} \wp^{(4)}(z,\tau )\right)  f(z,\tau) =0 \,,\\
 &\textstyle K_{1,1}=\frac{5}{2}\,,\quad K_{1,2}=\frac{7}{2} \,,\quad K_{1,3}=\frac{7}{2}  \,,\quad K_{1,4}=\frac{7}{2} \,,\quad K_{1,5}=\frac{7}{2} \,, \quad K_{1,6}=\frac{7}{2}\,,
\end{align}
where $f^{(n)}$ denotes the $n$th derivative in $z$ (also $'=\partial_z$). $E_r(\tau)$ is the Eisenstein series of weight $r$ 
and $\wp(z,\tau)$ is the Weierstrass $\wp$-function.
The differential equations for $k=1, 2, 3, 4$ are agree with those in \cite{MMS, MMS2}.


\end{document}